# Ultra-sensitive label-free in-situ detection of dynamically driven self-assembly of 2D-nanoplatelets on SOI chip

Benjamin T. Hogan[1,2], Sergey Dyakov[3], Lorcan J. Brennan[4], Salma Younesy[1,5], Tatiana Perova[6,7], Yurii K. Gun'ko[4,7], Monica F. Craciun[1] and Anna Baldycheva[1,6 *]

Fluid dispersed two-dimensional (2D) composite materials with dynamically tunable functional properties have recently emerged as a novel highly promising class of optoelectronic materials, opening up new routes not only for the emerging field of metamaterials but also to chip-scale multifunctional metadevices. However, *in-situ* monitoring and detection of the dynamic ordering of 2D nanoparticles on chip and during the device operation is still a huge challenge. Here we introduce a novel approach for on-chip, *in-situ* Raman characterisation of 2D-fluid composite materials incorporated into Si photonics chip. In this work the Raman signal for 2D nanoplatelets is selectively enhanced by Fabry-Pérot resonator design of CMOS photonic-compatible microfluidic channels. This has then been extended to demonstrate the first *in*-situ Raman detection of the dynamics of individual 2D nanoplatelets, within a microfluidic channel. Our work paves the way for the first practicable realisation of 3D photonic microstructure shaping based on 2D-fluid composites and CMOS photonics platform.

[1] Department of Engineering and Centre for Graphene Science, College of Engineering, Mathematics and Physical Sciences, University of Exeter, Exeter, EX4 4QF, UK.
[2] EPSRC Centre for Doctoral Training in Electromagnetic Metamaterials, University of Exeter, UK, EX4 4QL.
[3] Skolkovo Institute of Science and Technology, Photonics and Quantum Material Centre, Nobel street 3, Moscow, Russia.
[4] School of Chemistry and CRANN, University of Dublin, Trinity College, Dublin 2, Ireland.
[5] École Nationale Supérieure de Mécanique et des Microtechniques, Besançon, France.
[6] Department of Electronic and Electrical Engineering, Trinity College, The University of Dublin, Ireland.
[7] ITMO University, 49 Kronverskiy pr., St.-Petersburg, 197101, Russia.
*Corresponding author: a.baldycheva@exeter.ac.uk.

The recent development of nanocomposites of fluid-dispersed atomically thin two-dimensional (2D) materials has sparked a great level of interest as a promising *in-situ* tailored meta-material device platform for the next generation of multi-functional (opto)-electronic systems [1–13] with a wide range of important applications, such as renewable energy, optical communications, bio-chemical sensing, and security and defence technologies[8]. Dynamically controlled three-dimensional self-assembly of suspended 2D liquid exfoliated nano-flakes not only provides a breakthrough route for technological realization of 2D material based 3D device architectures[14–17], but also its fluidic nature allows CMOS-compatible integration on chip using microfluidic technology[18–20]. This opens up almost limitless possibilities in the fabrication of compact and low-power systems for the realisation of commercially viable, miniaturised, multi-functional light-management devices, for example light sources[21], tuneable optical filters[22] and nano-antenna phased arrays[23]. An example of a future device is shown in Fig.1, where dynamically reconfigurable 2D material fluid metastructures are integrated into a microfluidic system and coupled with a CMOS photonic circuit. Due to recent advances in CMOS photonics, the practicable application of any 2D material fluid composites, in particular those dispersed in liquid crystals (2D-LCs), in microfluidic electronic-photonic devices is technologically viable. The primary challenge for the first realization of on-chip controlled assembly of 2D flakes into functional micro-structures is a lack of a reliable and sensitive method for their *in-situ* characterisation that would provide information not only about integration of the composites in the device, but most importantly could enable to determine parameters such as spatial alignment and size of incorporated 2D nano-sheets in dynamics during the device operation. There are several existing characterisation methods which could be applied to overcome this challenge; including, Raman spectroscopy[5,24–29] and coherent anti-Stokes Raman spectroscopy (CARS)[30], as well as Fourier Transform Infrared Spectroscopy (FTIR)[31]. However, none of these techniques are suitable for investigation of fluid nanocomposites with relatively low concentrations of nanoparticles dispersed, since the significantly greater scattering volume always increases the intensity of the vibrational signal of the host fluid .Therefore, monitoring of the signal from weaker bands associated with dispersed nanoparticles is impossible. Furthermore, laser heating of liquids can cause their evaporation or even burning at high powers or after long exposure times, limiting the options for increasing signal-to-noise ratio. Most importantly, none of the existing characterisation methods can be used for *in-situ* characterization of fluid dispersed nanoparticles integrated into CMOS photonics chip. Here, we demonstrate for the first time a novel on-chip, *in-situ* micro-Raman characterization approach, whereby the Stokes Raman signal of 2D dispersed flakes is selectively enhanced at given wavelengths through the design of opto-fluidic channel geometry on silicon-on-insulator (SOI) platform. The developed characterisation approach demonstrates ultra-high signal sensitivity to the *xyz* alignment of 2D flakes within opto-fluidic waveguide channels during application of external stimuli. Our findings thus demonstrate for the first time the ability to monitor the dynamics of fluid-dispersed 2D nano-objects on chip.

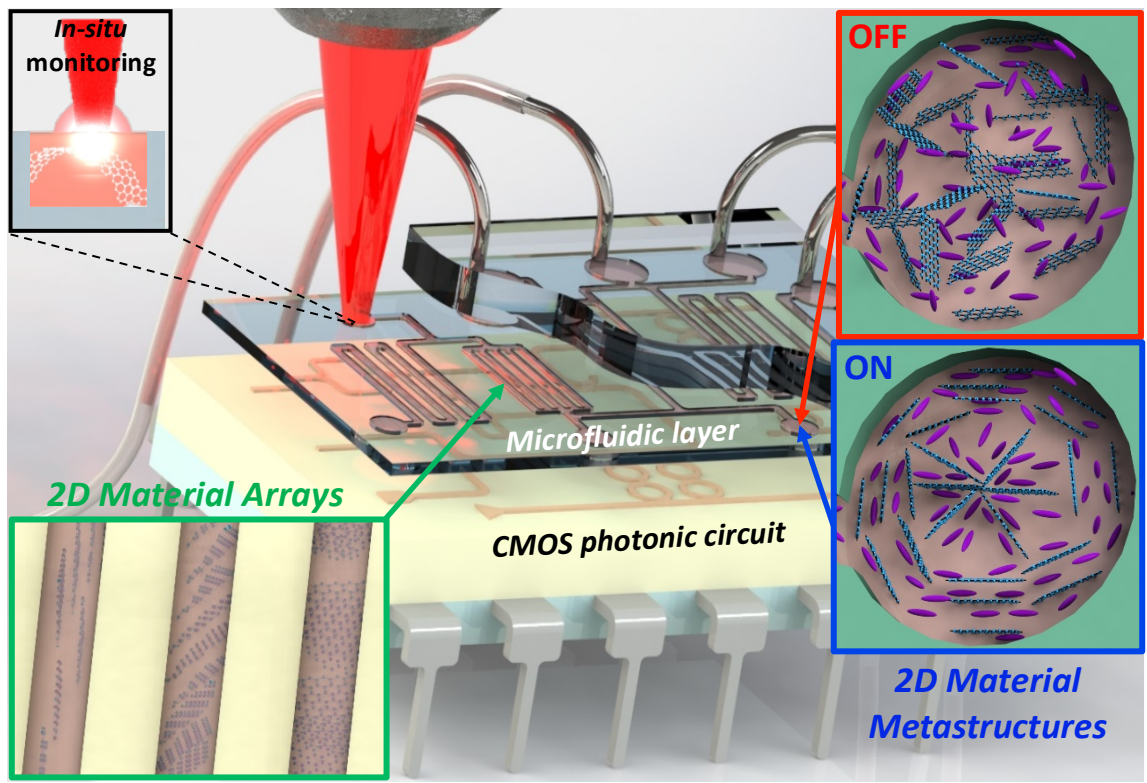

**Figure 1**: A future device based on a CMOS photonic circuit coupled to a microfluidic layer integrating dynamically reconfigurable 2D material metastructures, with *in-situ* micro-spectroscopy detection and monitoring.

***Enhanced Raman scattering through microfluidic structure design.*** Currently, large-scale CMOS photonics typically builds on a silicon-on-insulator (SOI) platform- a high-index contrast waveguide platform which prevents interference between the photonic integrated circuit components and the substrate. In this work, a microfluidic channel design on SOI platform (Fig. 2) is proposed to enhance the Raman signal intensity of the incorporated 2D nanomaterial-LC nanocomposite system. Infinitely thick silicon walls were considered; the thickness of the silicon walls have been shown to have little effect on the intensities in the simulations. The proposed Fabry-Pérot microfluidic cavity designs can therefore be readily integrated within a photonic device layer as a back-end process. i.e. without the need for fabrication of freestanding microstructures. Spontaneous Raman scattering is a quantum mechanical process with a random spatial distribution of the photons involved, however the optical behaviour of the scattered light can be modelled using a classical electrodynamics approach. Therefore, 2D flakes of LC nanocomposites can be modelled as a system of chaotically oriented oscillating electrical dipoles[26] within a microfluidic channel, with the dipole emission defining the Raman signal wavelength. To simulate the far-field intensity of the dipole emission we use the scattering matrix method [See Methods]. Since the incident and Raman scattered wavelength of light is considerably greater that the thickness of 2D flakes in nanocomposites, the flake is considered to be a point dipole- an emitter of only Stokes or anti-Stokes photons within the system- and hence the refractive index of the flake material has no effect on the propagation of light in the channel.

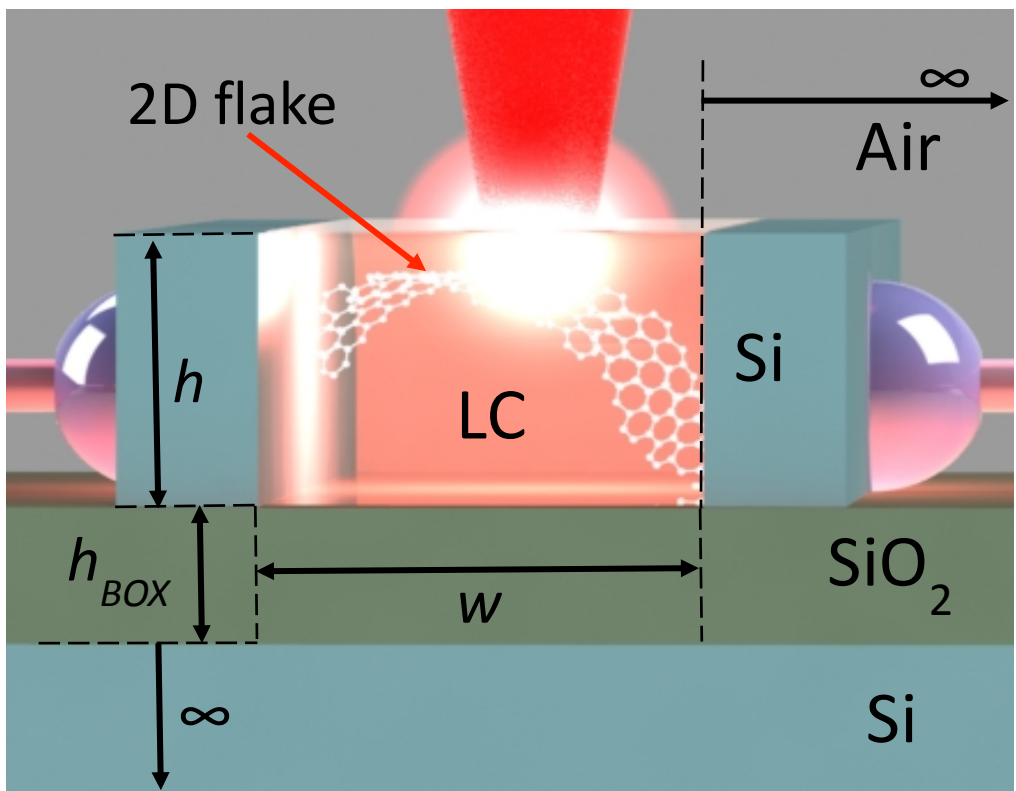

**Figure 2**: Schematic of the optofluidic waveguide structure based on 2D-fluid nanocomposites, on SOI platform, optimised for *in-situ* Raman measurements of individual 2D nanoplatelets.

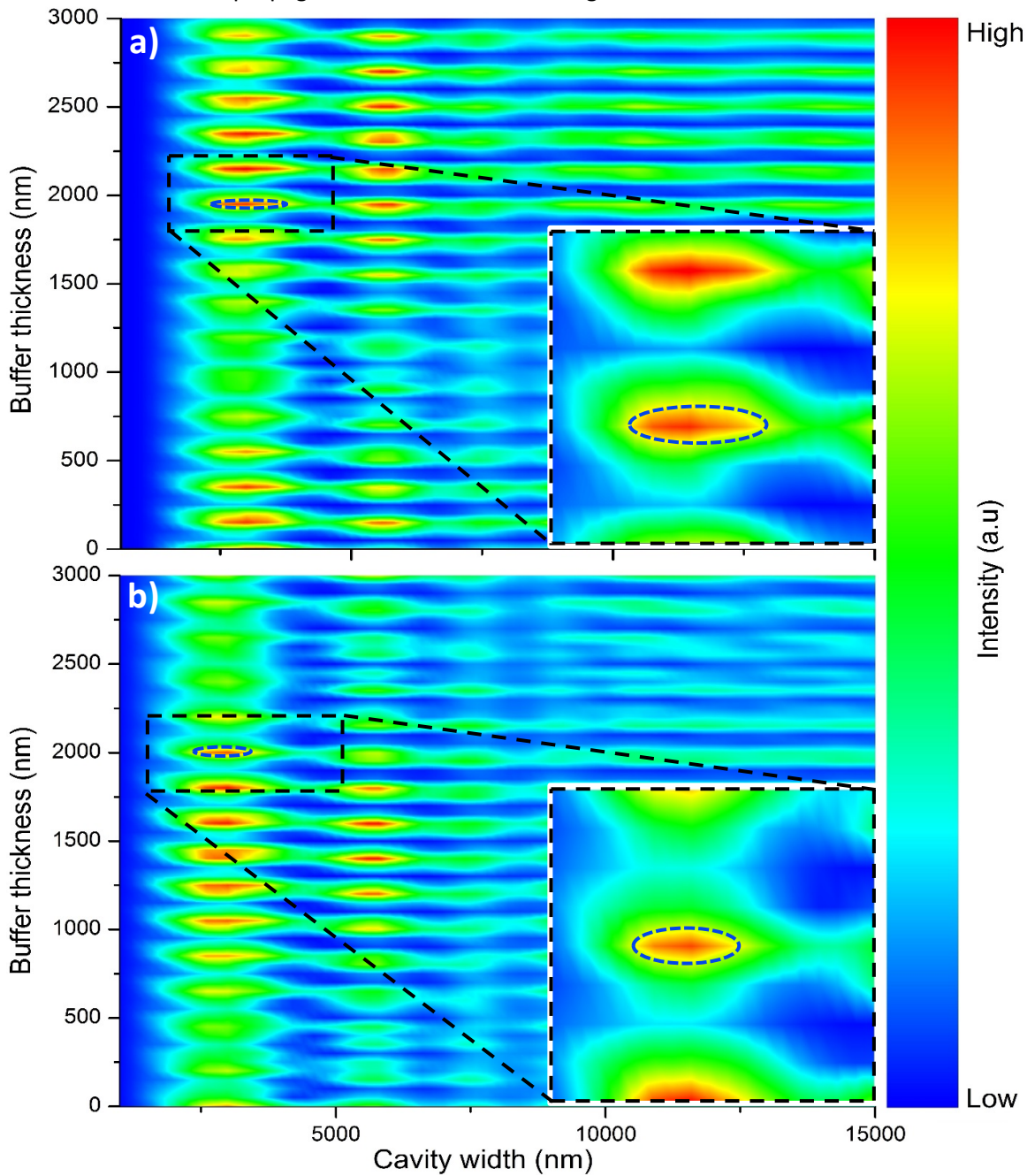

**Figure 3**: Maps of the simulated Raman signal intensity at wavelength corresponding to **a)** the D band and **b)** the G band of 2D carbon-based materials, under excitation by a 532nm Raman laser, as the microfluidic cavity width and buffer oxide layer thicknesses are varied. High intensity areas of interest are highlighted.

The Raman signal intensity variation is investigated across a range of values for two microfluidic channel parameters- the cavity width, $w$, and the buffer oxide (BOX) layer thickness, $h_{BOX}$. For Fabry-Pérot effects in a layer to be observed experimentally, its thickness should be less than the coherence length of the Raman scattered light[26]. However, this condition is not fulfilled for the silicon substrate layer, hence it is modelled as a semi-infinite material, by removing all Fabry-Pérot resonances within the substrate layer. The backscattered Raman signal intensity can be precisely predicted for wavelengths corresponding to the Raman active bands of a 2D-LC nanocomposite material using the SMM. For example, in this paper we consider the specific case of the *D* and *G* bands of graphene oxide (GO), dispersed in a nematic LC host. All simulations are made for normal angles of Raman laser incidence and signal collection. The electric field vector of the incident light is oriented parallel to the channel walls. The nematic LC host is a birefringent material, however, an advantage of the microfluidic infiltration into SOI microcavities is the spontaneously induced planar alignment of the LC through interaction with the surfaces[32]. The LC will therefore have a director which is either parallel or perpendicular to the walls of the channel such that only either the extraordinary or ordinary refractive index is required [See Methods].

To engineer the parameters giving maximal intensity of the Raman signal, maps are produced for specific vibrational bands corresponding to the D and G bands of 2D carbon-based materials (Fig. 3). The periodic nature of the Raman intensity, as the BOX layer thickness is varied, is expected due to Fabry-Pérot type resonances within the layer. The series of maxima and minima of the Raman intensity observed with the variation of the cavity width can also be rationalised as a series of Fabry-Pérot type resonances. As the cavity width decreases, the amplitude of the peaks in the Raman intensity is enhanced significantly.

In order to experimentally demonstrate the enhancement of the Raman intensity, 2D material-fluid nanocomposites consisting of GO nanoplatelets were dispersed in commercially available LCs which were synthesised by a liquid phase dispersion method[33] [See Methods]. The optimal parameters for the microfluidic structures were selected from the combined maps for both D and G bands simultaneously (Fig. 4a). Fabry-Pérot resonator devices consisting of microfluidic channels of different widths were fabricated on <100> p–type SOI wafer with a thick buffer oxide *($h_{BOX}$=2 µm)* layer- typical to photonic layers in the photonic integrated circuits and with a thick silicon device layer of 15 *µm* in order to establish practically the range of all possible flake alignments for dynamic reconfiguration applications. Electron-beam lithography followed by plasma etching was used to fabricate the microfluidic structures. The synthesised GO-LC nanocomposites were integrated into microfluidic channels through capillary action *via* the infiltration reservoirs (100 *µm* across) on the chip (Fig. 4b-h)[26,34]. The two channels fabricated on the chips had widths of 3.6 *µm* (narrow channel), to give strong enhancement, and 11.6 *µm* (wide channel), to give comparably weaker enhancement. Scanning electron microscopy (Fig. 4c-e) and optical microscopy (Fig. 4f-h) both confirmed the successful integration of the nanocomposite into both channels, with GO flakes clearly present in each.

We first selected a LC for which there are no strong bands overlapping with those for GO, that allows for clear determination of the GO bands in gathered spectra. MLC 6608 exhibits weak Raman bands that fulfil this criterion. Raman spectra are presented for the GO-LC nanocomposite, in Fig. 5a, at three points *in-situ* on the chip; more specifically: in the wide channel (Fig. 5b), in the narrow channel (Fig. 5c) and in the infiltration reservoir (Fig. 5d). For the GO D band, Raman intensities were observed in the approximate ratio *5:8:16* for the infiltration reservoir, wide and narrow channels respectively. Similarly, for the G band, intensities were observed in the ratio *5:7:16*.

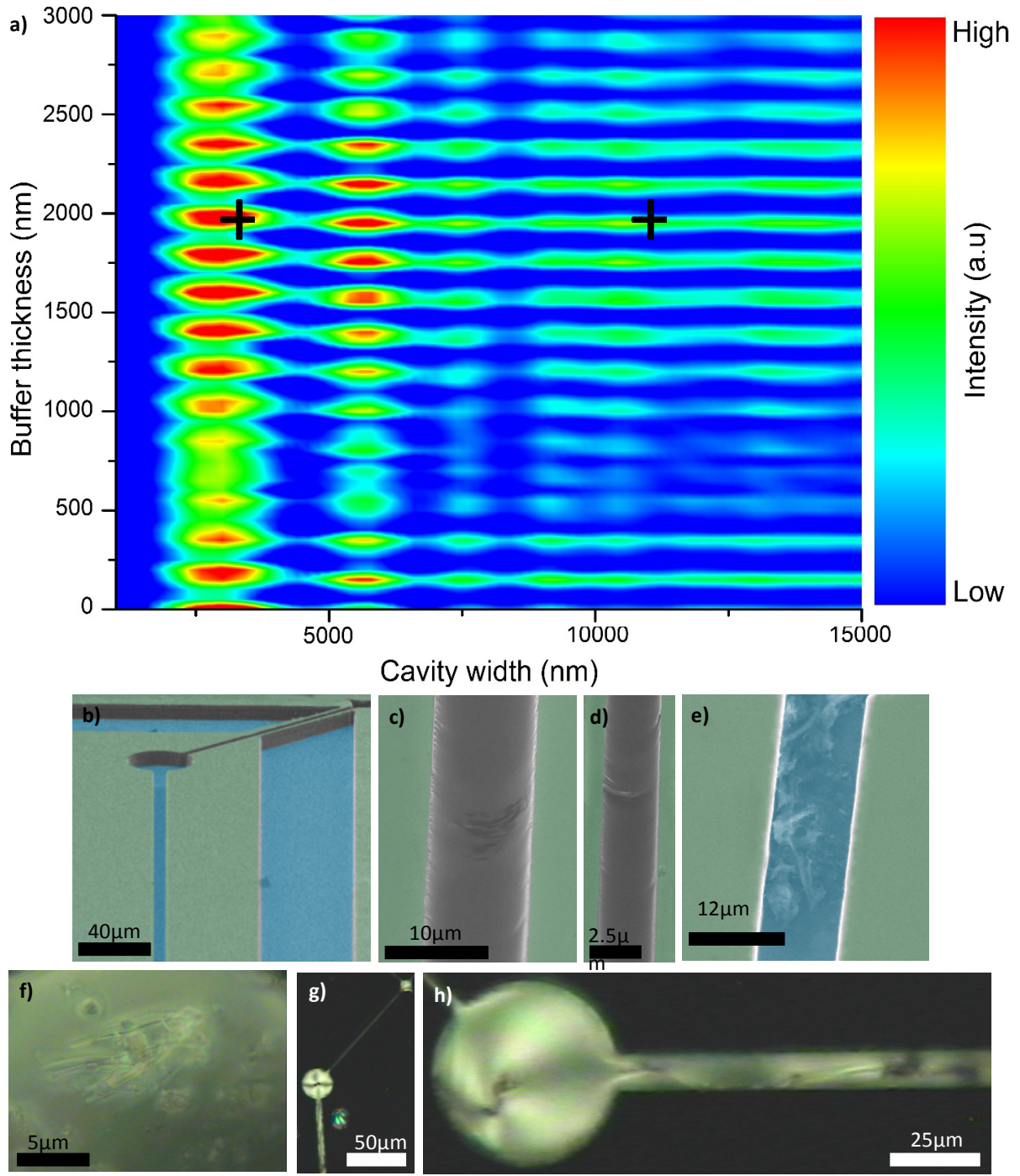

**Figure 4**: **a)** Map showing regions where both the of both the D and G bands of 2D carbon-based materials are predicted to be mutually enhanced in Raman spectroscopy measurements. Black crosses show the parameters of the channels fabricated. **b)** SEM image of the chip used for Raman measurements, before infiltration. **c)** SEM of GO flakes beneath the surface of infiltrated LC within an 11.6 $\mu m$ channel. **d)** SEM of GO flakes at the top surface of the LC in a 3.6 $\mu m$ channel. **e)** SEM of the 11.6 $\mu m$ channel, with the LC evaporated, showing the integration of the GO flakes. **f-h)** Polarised microscopy images of the structure infiltrated with a composite of MLC 6608 and graphene oxide. Integration of the composite into all microfluidic features on the chip can be seen. **f)** shows patterning of the LC surface within the reservoir, observed due to the presence of GO flakes.

The close agreement between the relative intensities of the D and G bands observed experimentally in the wide and narrow channels and those predicted by simulations (Fig. 5e) verifies the method for predicting the enhancement of the Raman signal intensities. Therefore, this technique presents an effective tool for designing structures which will maximise the enhancement of the Raman signal at given wavelengths. For both the D and G bands, the predicted enhancement ratio was slightly smaller (5%) than that which was observed experimentally. This difference is probably due to the flake not being positioned precisely at the centre of the channel in the experiment.

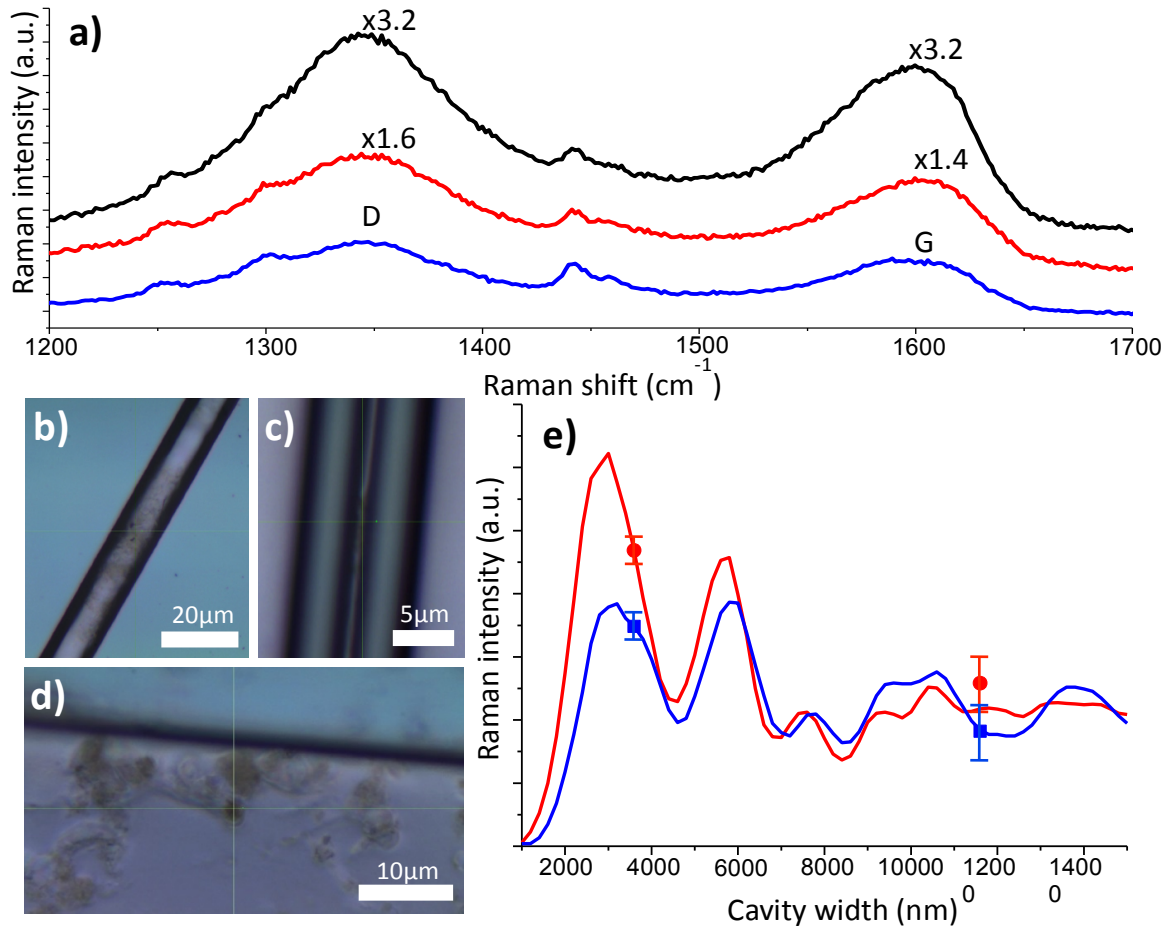

**Figure 5**: **a**) Raman spectra showing the enhancement of the D and G band bands for graphene oxide dispersed in liquid crystal MLC 6608 three microfluidic geometries: in **b**) an infiltration reservoir of width 100μm (blue), in **c**) a microfluidic cavity of width 11.6μm (red) and in **d**) a microfluidic cavity of width 3.6μm (black). **e**) Comparison of simulated (lines) and experimentally measured (points) Raman intensities of the graphene oxide D (blue) and G (red) bands. All data is normalised to the case where the walls are separated by a distance great enough for Fabry-Pérot resonances to have no effect.

***Real-time in-situ monitoring of 2D nanoplatelet alignment.*** Understanding of the dynamics of 2D nanoplatelet spatial alignment is essential for the first practicable realisation of three-dimensional metastructure formation on-chip. *In-situ* applied external stimuli to the microfluidic channel, such as applied electric field [6,15], magnetic field[35] or light coupling[36], induces dynamic re-ordering of suspended 2D flakes. More specifically, flakes can be induced to rise within the LC or to move towards (or away from) the walls of the channel. Here, a Raman laser was exploited to move the GO flake within the microfluidic channel (Fig. 6a), while at the same time being used to monitor the xyz alignment over time. The effect of the flake position on the Raman signal intensity was modelled by varying the position of the oscillating dipoles within the microfluidic channel. The liquid crystal in this case was chosen as commercially available E7 and the refractive index adjusted accordingly [see Methods]. We choose the optimal design parameters of the microfluidic channel that demonstrate strong enhancement of D and G bands, i.e. a narrow microfluidic channel with a width of 3.6 $\mu m$. E7 has a strong Raman active vibrational band at around 1605 $cm^{-1}$, overlapping with the G band. Nevertheless, utilising the proposed signal enhancement design, the observation of the G band is feasible, as demonstrated in Fig 6b. All other parameters remain the same as in the previous simulations, with the intensity predicted separately for variations in the lateral and vertical displacements. Vertical displacements are defined from the bottom surface of the channel and lateral displacements from a side wall. For vertical displacement variation, the flake is fixed at the lateral centre (1800 $nm$ [or $w$/2] from the wall) while for lateral displacement variation it is fixed at the vertical centre (7500 $nm$ [or $h$/2] from the bottom of the channel).

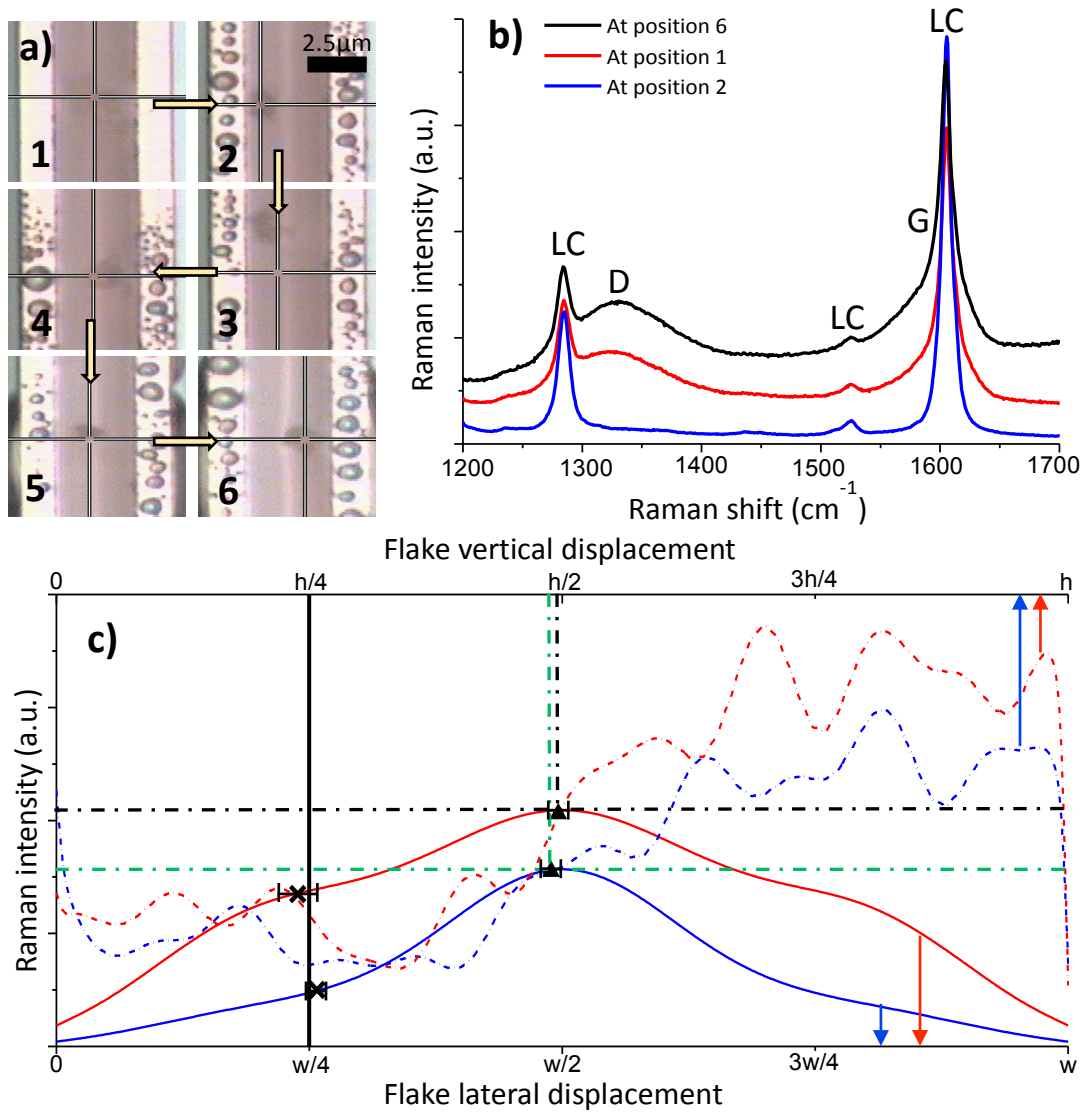

**Figure 6**: **a**) GO flake movement induced by the Raman laser. Each image represents the change after 10 $s$ exposure time in the order in which they were observed. **b**) Raman spectra of a GO flake dispersed in liquid crystal E7 within a narrow channel (approx. 3.6 $\mu m$) at positions 1 (red), 2 (blue) and 6 (black) as seen in **a**. **c**) Simulation of the variation of the Raman intensity of the GO D (blue) and G (red) bands for lateral (solid lines) and vertical (dashed) displacements. Black crosses correspond to the lateral displacement predicted from the normalised Raman intensity at point 6 in **a** while the black line represents the approximate displacement measured using optical microscopy. The horizontal green and black dashed lines represent the experimentally measured Raman intensities of the GO D and G bands respectively at point 5 in **a**. The vertical dashed lines show the vertical displacements corresponding to these values.

The ratio of the D and G band intensities is not constant as the position is varied. For lateral displacements, the strongest signal is observed at the centre of the channel in both cases, while the lowest signal is demonstrated at both edges of the channel (Fig. 6c). There are ratios of 11 and 32 between the minimal and maximal intensities predicted for the D and G bands respectively. For vertical displacements, there is a general increase in the average intensity as the vertical displacement is increased (Fig. 6c). The maximum and minimum values of the average intensity differ by factors of around four and five times for the D and G band wavelengths respectively.

Spectra for the same GO flake in multiple positions within a 3.6 *μm* microfluidic channel are presented in Fig. 6a-b. Very low signal corresponding to GO is recorded when the flake is positioned next to the wall of the channel. Greater intensity is recorded as the flake is moved towards the top surface of the LC within the channel (the relative vertical displacement is approximated from the apparent colour change observed in the composite at the position of the flake under optical microscopy).

For the lateral displacement of the flake, the ratios of the intensities of the GO D and G bands were extracted from the experimental spectra when the flake was next to the wall and when moved further towards the centre of the channel. Ratios of 13±1 and 7.9±0.4 are found for the D and G bands respectively. These ratios were then used as a multiplier on the simulated intensity for the flake next to the wall to determine a predicted displacement. Values of 1180±50 *nm* and 1100±100 *nm* are predicted from the D and G bands respectively, which corresponds closely with the value determined from optical microscopy of 1150±100 *nm*.

For the vertical displacement of the flake, again the ratios of the intensities of the GO D and G bands were extracted, this time from data with the flake at the bottom of the channel and further towards the surface. The ratios were then used to multiply the simulated intensity with the flake at the bottom of the channel to determine a predicted displacement. Values of around 7300±150 *nm* and 7440±60 *nm* were predicted from the D and G bands respectively. It should be noted that some intensities would give multiple possibilities for the predicted position of the flake.

For the lateral flake displacement, predictions made using experimental data in tandem with the simulated results are closely matched to the value measured by optical microscopy techniques, falling within the experimental error of that measurement. The disagreement between the values predicted using the ratios of the D and G bands is potentially caused by the drift of the flake during measurements. The lateral position can be determined with similar precision to optical microscopy. Reducing the signal-noise ratio in the experimental spectra would reduce the error in the predicted values.

Similarly, the simulations covering the effect of vertical flake position also agree closely with the experimental data. The vertical position of the flake cannot be determined from optical microscopy as there is no reference point to compare to for either direct measurement or optical contrast techniques, however, the close agreement of the positions determined separately from the D and G bands suggests that the method is accurate. Therefore, the predictions made from the Raman spectra are the most accurate method of determining the vertical position currently available. This provides the first step towards a method for the *in-situ* monitoring of the orientation, position and alignment of flakes within GLC composites. Again, the disagreement between the values predicted from the D and G bands may be due to the drift of the flake. Determination of the flake position in co-ordinates parallel and orthogonal to the incident light is therefore readily achievable, given a suitable spectrum at a known reference point.

The data presented covers the effect of two structural parameters, however including the channel depth as a further parameter would provide an additional dimension to the control over the selectivity of the signal enhancement. Additionally, further control could be gained by the variation of the materials used for the buffer layer and the walls of the channel to modify the permittivities of the different layers. This represents a route towards designing structures to selectively enhance Raman signal from specific vibrational bands of a nanomaterial.

For materials with weak characteristic Raman bands, selective wavelength enhancement opens up the possibility of increasing the ease of observation of the weak Raman bands, even in the presence of stronger Raman bands. This is particularly useful for the characterisation of composite materials where the Raman bands of components with much smaller scattering volumes could be selectively enhanced. Signal enhancement means that lower laser powers are required to give the same intensity response. This could be particularly useful for temperature sensitive materials where laser heating can be a significant concern. The principles determining the selectivity of the enhancement of the Raman signal also allow the same approach to be applied to anti-Stokes Raman bands.

In summary, we have developed a novel approach for on-chip, *in-situ* characterisation of GLCs, whereby the Raman signal can be selectively enhanced at given wavelengths by Fabry-Pérot resonator design of microfluidic channels. Enhancement of the signal has been predicted *via* simulation and confirmed from experimental spectra. The effect of nanoparticle position on the signal intensity has been considered *via* simulation and also observed in experimental spectra. It is expected that these methods will be used to determine the position of single particles within a composite.

## Methods.

**Scattering matrix method (SMM).** The scattering matrix method is a powerful tool for simulating of the near- and far-field light distribution for structures which can be split into layers uniform along at least one direction[37–40]. The main principle of this method is the decomposition of electric and magnetic fields into Fourier series in each layer and connecting the Fourier components of adjacent layers in accordance with the boundary conditions of Maxwell's equations. In this work, to reach convergence, 400 Fourier harmonics were used. The local components of the electromagnetic field were found, forming material matrices in each layer. By applying an iterative procedure, the total scattering matrix for the whole structure was calculated[41]. The back-scattered Raman intensity was calculated from the components of the scattering matrix. The spot size of the Raman laser was effectively considered as equal to the channel width. To account for this, all Raman intensities were normalised by accounting for the incident field strength in the channel.

**Calculation of LC refractive indices.** The ordinary and extraordinary refractive indices were calculated at all wavelengths used in the simulation using a three-coefficient Cauchy model, $n_{o,e} = A_{o,e} + \frac{B_{o,e}}{\lambda^2} + \frac{C_{o,e}}{\lambda^4}$; where $n$ is the refractive index and $A$, $B$ and $C$ are suitable fitting parameters. The subscripts $o$ and $e$ represent the ordinary and extraordinary indices respectively. For example, for commercially available LC MLC 6608, $A_o$, $B_o$ and $C_o$ were set to 1.4609, 5x10$^{-3}$ $\mu m^2$ and 0 $\mu m^4$ respectively to give good agreement with experimental measurements in the visible region[42]. As MLC 6608 has a low birefringence in the visible region, the two-coefficient Cauchy model is a sufficient approximation.

**Synthesis of GO-LC nanocomposites.** Graphene oxide (GO) was prepared from bulk graphite *via* the Hummers method[43]. 1 *g* of graphite, 0.5 *g* of sodium nitrate and 23 *mL* of sulphuric acid (H$_2$SO$_4$) were added to a 500 *mL* round bottomed flask and stirred at 4 *°C* for 15 minutes. 3 *g* of potassium permanganate (KMnO$_4$) was added slowly with vigorous stirring. Once all the KMnO$_4$ was added, the ice bath was removed and the suspension was heated to 35 *°C* for 30 minutes. This produced a murky brownish-grey solution. Following this, 46 *mL* of water was added and the suspension was set to stir for 15 minutes. The solution was then treated with 1.4 *mL* of hydrogen peroxide (H$_2$O$_2$). The product was washed through centrifugation up to 10 times with a 10 % aqueous solution of hydrochloric acid (HCl) followed by copious amounts of deionised water. The resulting GO was suspended in H$_2$O and filtered under vacuum onto an omniporous 200 *nm* membrane and washed with 1 *L* of H$_2$O. The resulting GO was then dried at 80 *°C* for several days. Once dried the GO was dispersed in tetrahydrofuran (THF) through ultrasonication. The dispersions were then centrifuged, allowing for extraction of the lowest mass GO flakes from the top layer of the suspension. An aliquot from this layer was mixed with the liquid crystal and the mixture underwent further ultrasonication to ensure thorough mixing of the two components. The resulting dispersion was dried in a Schlenk flask, under vacuum, allowing for complete evaporation of the residual solvent, resulting in a nanocomposite of GO nanoplatelets uniformly dispersed in the liquid crystal. The concentration of GO in the final nanocomposites was approximately 0.01 $g.ml^{-1}$.

**Instrumentation.** Scanning electron microscopy (SEM) measurements were performed on a Hitachi S3200N system with a practical operational magnification between 20-60000x, accelerating voltages from 0.3 to 30 *kV*, vacuum chamber pressure <0.1 *mbar* and a maximum resolution of 3.5 *nm*. Micro-Raman measurements were performed using a Renishaw 1000 system (with a 514.5 nm excitation wavelength from Ar$^+$ laser, a power of 5 *mW* and a spot size of approximately 3 *µm*) and on a Horiba Raman system (with an excitation wavelength of 532 *nm*, a power of 8.75 *mW* and a spot size of approximately 5 *µm* when focussed through a x50 objective). Polarised light images were obtained using a Zeiss Axioscope 2 microscope with a Zeiss Axiocam MRc 5 camera, with x20 and x50 objectives used.

## Acknowledgments

We acknowledge financial support from: the Engineering and Physical Sciences Research Council (EPSRC) of the United Kingdom via the EPSRC Centre for Doctoral Training in Electromagnetic Metamaterials (Grant No. EP/L015331/1) and also via Grants No. EP/G036101/1, EP/M002438/1, and EP/M001024/1, Science Foundation Ireland Grant No. 12/IA/1300, the Ministry of Education and Science of the Russian Federation (Grant No. 14.B25.31.0002) and the Royal Society International Exchange Grant 2015/R3. The microfluidic structures were fabricated at Tyndall National Institute under the Science Foundation Ireland NAP368 and NAP94 programmes.

## Author Contributions.

B.H. and A.B. conceived and planned the research, analysed data and wrote the paper with input from M.C., T.P. and Y.G.; B.H. and S.Y. conducted the experiments; S.D. wrote the code used for simulations; L.B. synthesised the nanocomposites used; A.B. and M.C. supervised and directed the research. All authors read and approved the final manuscript.